 \documentclass[final,5p,times,twocolumn,authoryear]{elsarticle}

\usepackage{amssymb}
\usepackage{amsmath}  
\usepackage{lipsum}
\usepackage{booktabs}
\usepackage{caption}    

\usepackage{multirow}
\usepackage{graphicx}
\usepackage{array}
\usepackage{booktabs}  
\usepackage[table,xcdraw]{xcolor}  
\usepackage{tcolorbox} 

\journal{Speech Communication}

\begin{document}

\begin{frontmatter}

\title{Heterogeneous bimodal attention fusion for speech emotion recognition}


\author[first]{Jiachen Luo}
\author[second]{Huy Phan}
\author[first]{Lin Wang}
\author[first]{Joshua Reiss*}

\affiliation[first]{organization={The Centre for Digital Music, Queen Mary University of London}
         } 
\affiliation[second]{organization={Reality Labs, Meta}
            }

\begin{abstract}
Multi-modal emotion recognition in conversations is a challenging problem due to the complex and complementary interactions between different modalities. Audio and textual cues are particularly important for understanding emotions from a human perspective. Most existing studies focus on exploring interactions between audio and text modalities at the same representation level. However, a critical issue is often overlooked: the heterogeneous modality gap between low-level audio representations and high-level text representations. To address this problem, we propose a novel framework called Heterogeneous Bimodal Attention Fusion (HBAF) for multi-level multi-modal interaction in conversational emotion recognition. The proposed method comprises three key modules: the uni-modal representation module, the multi-modal fusion module, and the inter-modal contrastive learning module. The uni-modal representation module incorporates contextual content into low-level audio representations to bridge the heterogeneous multi-modal gap, enabling more effective fusion. The multi-modal fusion module uses dynamic bimodal attention and a dynamic gating mechanism to filter incorrect cross-modal relationships and fully exploit both intra-modal and inter-modal interactions. Finally, the inter-modal contrastive learning module captures complex absolute and relative interactions between audio and text modalities. Experiments on the MELD and IEMOCAP datasets demonstrate that the proposed HBAF method outperforms existing state-of-the-art baselines.

\end{abstract}



\begin{keyword}
attention  \sep multi-modal sentiment analysis \sep conversational emotion recognition \sep intra-modal interaction  \sep inter-modal interaction 


\end{keyword}

\end{frontmatter}

\section{Introduction}
Conversational emotion recognition aims to understand human emotions and sentiment when interacting with one another during conversations and classify each utterance into its associated emotional state. Emotion recognition in conversations has found widespread applications in many tasks \citep{singh2022systematic, cai2023emotion, ezzameli2023emotion, geetha2024multimodal}. However, automatic recognition of emotions is a challenging problem in real-world scenarios as human emotional expressions are often diverse in nature across individuals and cultures \citep{singh2023universality, gong2023cross}.

\begin{figure*}[!t]
\centering
\includegraphics[width=1\linewidth]{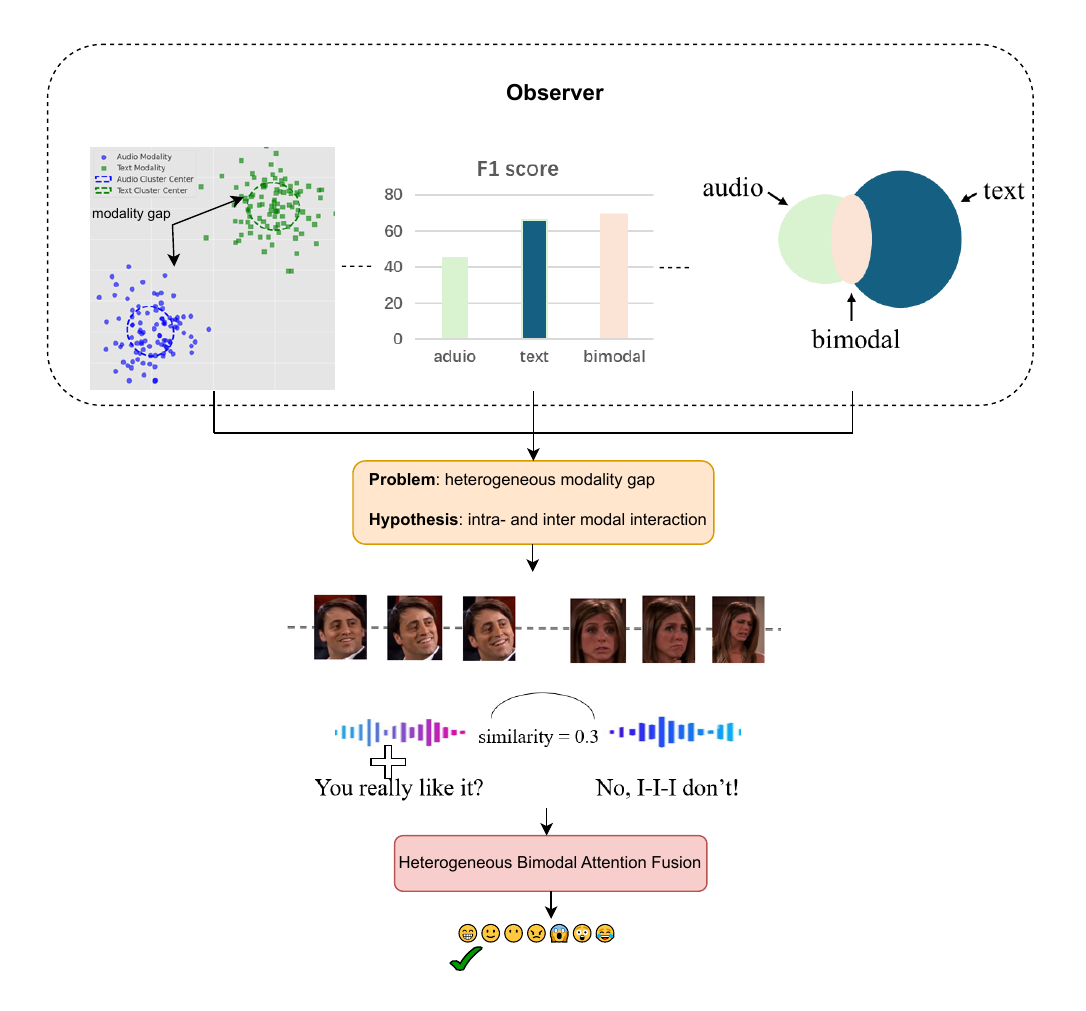}
\caption{Illustration of the proposed Heterogeneous Bimodal Attention Fusion (HBAF) method. Based on experimental observations, this figure illustrates key findings: (1) the presence of a modality gap between audio and text modalities, (2) the differing contributions of audio and text in bimodal speech emotion recognition, and (3) the existence of both unique and shared information between audio and text modalities. These observations lead to the identified problem of the heterogeneous modality gap and the hypothesis of leveraging intra- and inter-modal interactions. To address this, we propose the HBAF method for bimodal speech emotion recognition. The `similarity’ in the figure represents the alignment or similarity between audio and text modalities.}
\end{figure*}

Humans convey emotions through various modalities, including speech, facial expressions, and body postures \citep{ezzameli2023emotion, zhang2023learning}. Among these, speech stands out as a predominant, contact-free channel through which voice characteristics and linguistic content can be effectively communicated \citep{sebe2005multimodal, kessous2010multimodal}. It has been emphasized that an ideal model for conversational emotion recognition should integrate multi-modal information, as this approach better reflects human sensory systems \citep{calvert2004multisensory, turk2014multimodal}. For example, as shown in Figure 1, the short sentence `You really like it?' is ambiguous and can convey positive, neutral, or negative emotions depending on the context. It is challenging to determine the associated emotion based solely textual utterance. However, when accompanied by low voices or sobbing sounds, the sentiment behind the short sentence becomes easier to interpret as negative. Different modalities carry complementary information, and relying on a single modality is insufficient for accurate emotion recognition in real-world scenarios \citep{soleymani2011multimodal, liu2021comparing, zhang2024deep}. Therefore, combining information from multiple modalities provides richer, emotion-relevant insights by allowing modalities to complement or augment one another.

The heterogeneous gap and heterogeneous conflicts are key challenges in multi-modal emotion recognition \citep{cai2023emotion, geetha2024multimodal}. The heterogeneous gap refers to differences in representation levels between modalities, such as the abstract and contextual nature of text features compared to the detailed, low-level patterns in audio features. These differences make aligning and integrating information from multiple modalities difficult, as they require careful consideration of how each modality's characteristics complement or contrast with others. Heterogeneous conflicts arise when features from different modalities are directly fused into a shared representation space without proper alignment, leading to noise and misinterpretation. Addressing these issues is critical for building effective models that can leverage the complementary strengths of multiple modalities to reliably recognize emotions.

While previous methods have shown promise \citep{cai2023emotion, geetha2024multimodal}, significant challenges remain, particularly in learning from multi-modal data for conversational emotion recognition \citep{ezzameli2023emotion}. Addressing the heterogeneous gap and heterogeneous conflicts requires developing effective representations for each modality that can adapt to the variability of emotional expression across individuals. Moreover, capturing complex intra- and inter-modal interactions is critical for accurately recognizing emotional content. Contextual cues, especially in multi-speaker conversations, are essential and demand learning interactions across both low-level and high-level representations. However, most existing approaches focus on interactions at the same representation level \citep{tsai2019multimodal, liu2024contrastive}, overlooking the benefits of cross-level fusion. To overcome these limitations, we propose a novel audio context module that enriches low-level audio features with contextual information, enabling better alignment and interaction across modalities. By addressing the heterogeneous gap and conflicts, this approach provides a robust foundation for more effective and accurate multi-modal emotion recognition.

Building on efforts to address the heterogeneous gap and conflicts, a key challenge is the effective modeling of dynamic intra- and inter-modal interactions across modalities with varying contributions \citep{hazarika2018conversational, majumder2019dialoguernn, ezzameli2023emotion}. These interactions are crucial for bridging representation gaps and ensuring seamless integration of multi-modal data. However, many previous methods simply concatenate data at the input level, neglecting intra-modal relationships and amplifying noise. This approach often assumes equal contributions from all modalities, which rarely reflects reality. Advancements such as HGraph-CL have attempted to address these issues through hierarchical graph contrastive learning \citep{lin2022modeling}, but they struggle to disentangle modality-specific features and capture shared patterns necessary for effective inter-modal integration. We propose that inter-modal contrastive learning offers significant potential to overcome these limitations, enabling better alignment, integration, and utilization of complementary features to improve the robustness and accuracy of multi-modal emotion recognition.

Motivated by the above observations, we propose a Heterogeneous Bimodal Attention Fusion (HBAF) method that incorporates low-level audio representations to assist high-level text representations for bimodal emotion recognition. As shown in Figure 2, the proposed method mainly consists of three modules, uni-modal representation, multi-modal fusion, and inter-modal contrastive learning. 

In daily communication, a person might say, "I'm fine," with a sharp and irritated tone. While the text alone suggests a neutral or positive sentiment, the audio reveals frustration through tone, pitch, and speech rate. The HBAF method addresses this mismatch by enriching audio features with contextual information and dynamically adjusting attention to emphasize emotional cues from the audio while integrating the semantics of the text. This approach allows the model to accurately identify the underlying frustration, effectively resolving the heterogeneous gap in real-world scenarios. 
The main novelty of the proposed method is summarized as follows. 

First, we incorporate contextual information into low-level audio representations to project all modalities into a common shared subspace, aligning their contextual distribution to facilitate better fusion of heterogeneous modalities. Second, we design a multi-modal fusion module to discover comprehensive intra- and inter-modal interactions among different utterances across audio and text. Specifically, we rely on bimodal attention network, dynamic filter gate, and residual connection to train a multi-modal fusion network, enabling dynamic self- and cross-modal attentive weights to filter incorrect cross-modal interactions and exploit intra- and inter-modal relationships. Finally, we employ inter-modal contrastive learning module to model absolute and relative inter-modal content, encouraging the model to detect more complementary interactions between audio and text pairs. 

We validate the proposed method experimentally and  demonstrate that the paradigm for multi-modal emotion recognition primarily relies on supervised learning as the main method, with self-supervised learning serving as a complementary approach to enhance effectiveness and achieve a more comprehensive understanding of emotions. 





The rest of this paper is organized as follows. Section 2 presents a brief literature review. Section 3 describes our method in detail. Section 4 illustrates experiments. Section 5 demonstrates results and discussion. Finally, Section 6 gives conclusions based on this work.

\section{Related Works}
\label{sec:format}

\subsection{Feature Representation}
The quality of different modality information plays a decisive role in emotion recognition in conversations. To date, many studies have explored effective and efficient features in audio, textual, and visual modalities for multi-modal emotion recognition \citep{calvo2010affect, singh2022systematic, cai2023emotion,  ezzameli2023emotion}. Generally, audio feature engineering methods are roughly divided into two categories: prosodic and spectral features \citep{koolagudi2012emotion, wani2021comprehensive, hashem2023speech}. Most of multi-modal approaches use prosodic or hybrid features combining of prosodic and spectral features with traditional classifiers such as Hidden Markov Model \citep{schuller2003hidden}, and Support Vector Machines \citep{jain2020speech}. 

By contrast, due to the powerful representation capabilities of deep learning, some pre-trained audio models can automatically explore more informative features \citep{liu2022audio, zaman2023survey}. Pre-trained audio models, such as, VGGish \citep{gemmeke2017audio}, openL3 \citep{cramer2019look}, wav2vec\citep{baevski2020wav2vec}, data2vec \citep{baevski2022data2vec}, etc., have accelerated the development of automatic speech recognition, voice conversation, and emotion recognition. For example, Hung {et al.} employed OpenL3 feature-informed embedding space regulrarization for audio classification \citep{hung2019multitask}. Eunjeong et al. investigated the effectiveness of L3-Net and VGGish deep audio embedding methods for music emotion inference over four music datasets \citep{eunjeong2021music}. Pepino {et al.} have also explored the use of the wav2vec 2.0 model as a feature extractor for emotion recognition \citep{pepino2021emotion}. 

For textual representation, much process has been made in learning embedding of individual words such as word2vec \citep{mikolov2013distributed}, GloVe \citep{pennington2014glove} and of phrases and sentences such as doc2vec \citep{le2014distributed}. More recently, there exists growing attention over the development of large unsupervised pre-trained language models \citep{wang2023pre}. Pre-trained language models such as ELMo \citep{peters2017semi}, GPT \citep{radford2018gpt} and BERT \citep{devlin2019bert}, have achieved high performance in various tasks by constructing contextual representation \citep{deng2021survey,al2024challenges}.  By pretraining on large-scale unsupervised texts, these models enable learning linguistic representation related to the emotional states. In \citep{acheampong2021transformer}, Acheampong {et al.} analyzed the efficacy of BERT, RoBERTa, DistilBERT, and XLNet pre-trained transformer models in recognizing emotions from text. 

Commonsense knowledge is essential for modeling the structure and flow of the dialogue, as well as the emotional dynamics of the participants \citep{zhou2018commonsense, lin2020ket}. Commonsense knowledge is beneficial to emotion understanding. The commonsense transformer model COMET is commonly used to extract the commonsense features \citep{bosselut2019comet}. A typical example in this line of work is Deepanway {et al.}, who  incorporated different elements of commonsense such as intra- and inter-speaker information and causal relations, and builds upon them to learn interactions between interlocutors participating for emotion identification in conversations \citep{ghosal2020cosmic}.

\subsection{Multi-modal Emotion Recognition}
After feature extraction, there are two main multi-modal fusion strategies: feature fusion \citep{majumder2019dialoguernn, chudasama2022m2fnet} and decision fusion \citep{poria2017review, zhao2021multimodal}. Poria {et al.} \citep{poria2018meld} and Zhou {et al.} \citep{zhou2018inferring} leverage multi-modal information through concatenating features from different modalities without modeling the interaction between modalities. Lee {et al.} proposed weighted sum and weighted product rules for audio, text and video decision-level fusion, avoiding the difficulty of fusing heterogeneous information \citep{lee2021multimodal}. But it is unable to capture interactions of different modalities. 

To get better emotion recognition information, conversational emotion recognition requires deep understanding of human interactions in conversations \citep{singh2022systematic, cai2023emotion,geetha2024multimodal}. However, there is no provision to model interactive influences. Additionally, conversational emotion recognition primarily focuses on employing deep-learning based algorithms to carry out contextual modeling within either a textual or multi-modal setting \citep{mittal2020emoticon, deng2021survey}. Numerous studies have emphasized that emotional dynamics are interactive in nature, rather than confined to individuals and following a unidirectional pattern \citep{calvo2010affect, singh2022systematic,cai2023emotion}. These studies have attempted to capture such dynamics by examining transition properties. 

Recently, several studies employ memory networks to capture the intra- and inter-modal interactions in two-speaker conversations \citep{hazarika2018icon, wang2024dynamic}. Majumder {et al.} proposed a RNN-based neural architecture that kept track of the intra-modal interaction throughout the conversation and used this information for emotion classification \citep{majumder2019dialoguernn}. Hazarika {et al.} took the multi-modal approach comprising audio, visual and textual features with greater recurrent units to model past utterances of each speaker into memories. Such memories were then merged using attention-based hops to capture inter-speaker dependencies \citep{hazarika2018conversational}. Hazarika {et al.} designed interactive conversational memory network, a multi-modal emotion detection framework that extracted multi-modal features from conversational videos and hierarchically modeled the intra- and inter-speaker emotion influences into global memories. Such memories generated contextual summaries which aided in predicting the emotional orientation of utterance-videos \citep{hazarika2018conversational}. 

Besides, Ren {et al.} introduced a cross-modal attention fusion module to capture cross-modal interactions of multi-modal information, and employed a conversational modeling module to explore the context information and speaker dependency of the whole conversation. Concretely, the cross-modal attention fusion module captured the cross-modal interactions and complementary information among the pre-extracted uni-modal features from acoustic, textual and visual modalities based on the cross-modal attention block. Afterward, the updated features from each modality were fused to concentrate more on the informative modality and achieve a refined feature for each constituent utterance. The conversational modeling module defined three different gated recurrent units with respect to the context information, the speaker dependency, and the emotional state of utterances \citep{ren2021interactive}. However, previous work did not pay significant attention to long-term contextual information, speaker-sensitive content and intent information \citep{hazarika2018conversational, majumder2019dialoguernn}. Due to its effectiveness, the transformer attracted some in the field for its ability to combine multi-modal features.

Xie {et al.} proposed a Transformer-based cross-modality fusion with the EmbraceNet architecture to combine all the representation vectors from each modality \citep{xie2021robust}. Li {et al.} designed a new structure named Emoformer to extract multi-modal emotion vectors from different modalities and fuse them with sentence vector to be an emotion capsule. Furthermore, they designed an end-to-end conversational emotion recognition model which extracted emotion vectors through the Emoformer structure and obtain the emotion classification results from a context analysis model \citep{li2022emocaps}. Chudasama {et al.} designed a multi-modal fusion network that extracted emotion-relevant features from visual, audio and text modality \citep{chudasama2022m2fnet}. It employed a multi-head attention-based fusion mechanism to combine emotion-rich latent representations of the input data. But it ignored the mismatch between different modalities. 

\begin{figure*}[!t]
\centering
\includegraphics[width=1.0
\linewidth]{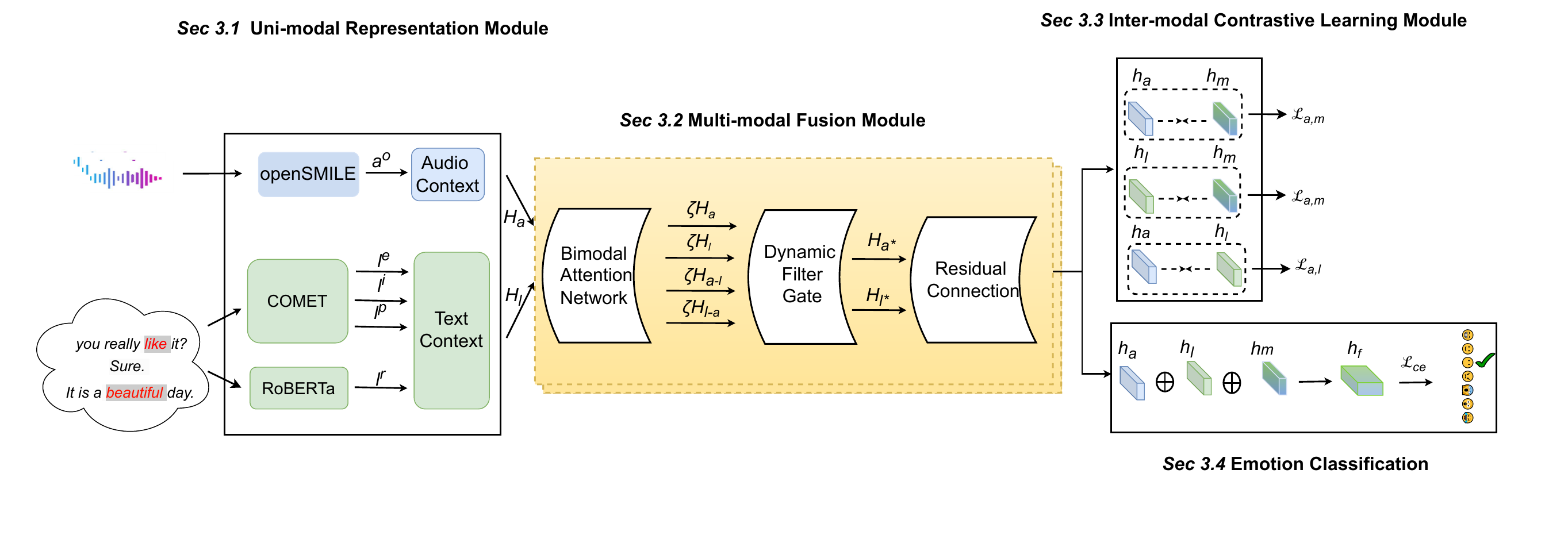}
\caption{Block diagrams of the proposed heterogeneous bimodal speech emotion recognition method (HBAF). The method consists of three modules, uni-modal representation, multi-modal fusion, and inter-modal contrast learning, with the details depicted in Figures 3-5, respectively. }
\label{fig:res}
\end{figure*}

Existing works on bimodal speech emotion recognition share some common limitations. First, most approaches do not adequately account for low-level audio representations, which can serve as important indicators of emotion \citep{wani2021comprehensive, singh2022systematic}. Second, since text tends to carry more information than audio \citep{calvo2010affect, cai2023emotion}, many approaches emphasize the text while neglecting the relationship between audio and text. Third, previous works fail to fully leverage the associations across different levels of multiple modalities, limiting their ability to model mutual correlations and capture long-term contextual dependencies \citep{ren2023maln}. Unfortunately, the existing approaches can not guarantee the cross-modality interaction at different levels, and often the learning of intra- and inter-modal will lose some semantic content. To address this challenge, we propose HBAF method to dynamically capture intra- and inter-modal interactions between low-level audio representation and high-level text representation. 

\section{Methodology}
\label{ssec:subhead}
Our goal is to infer the emotion of utterances presented in multi-turn and multi-speaker conversations. Emotion recognition in conversations task includes \emph{C} emotion categories, whose set is \emph{E} = {\emph{E}$_1$, \emph{E}$_2$, ..., \emph{E}$_C$}. Let us define a dialogue with $N$ utterances \emph{u}$^1_m$, \emph{u}$^2_m$, \ldots, \emph{u}$^N_m$, $m \in \{a, l\}$, where \emph{a}, \emph{l} represent audio and textual modality, respectively. For each utterance \emph{u}$^n_m$, $1\le n \le N$, an emotion label \emph{E}$_c$, $1\le c \le C$, is assigned.
\\
As shown in Fig. 2, the proposed method HBAF is composed of three key modules: uni-modal representation module, multi-modal fusion module, and inter-modal contrastive learning module. We provide a detailed explanation of each component below.

\subsection{Uni-modal Representation Module}
\subsubsection{Audio Features Extraction and Context Network} 

\textbf{Audio Features Extraction:} For the audio modality, we use the openSMILE toolkit to extract 6373 dimensional features constituting several low-level descriptors and various statistical functionals of varied vocal and prosodic features ($a^o$) \citep{eyben2010opensmile}. Following the baseline work \citep{poria2018meld}, we address the high dimensionality of the audio representation by employing SVMs for feature selection, resulting in a dense representation of the overall audio segment with a dimension of 512.

\textbf{Audio Context Network:} 
To bridge the heterogeneous modality gap, we employ an audio contextual network to enhance low-level audio representations ($H_a$) with a dimension of 512. These low-level descriptors (LLDs), which include statistical features such as pitch, energy, and prosodic functionals, are lack temporal or sequential information. This makes them insufficient for capturing the contextual emotional patterns present in audio data. To address this limitation, the audio contextual network introduces contextualized audio information by modeling relationships across the entire audio sequence. This transformation enriches the LLDs with temporal and sequential context, making them more effective for emotion understanding.

The audio contextual network consists of a convolutional layer, two bidirectional long short-term memory (LSTM) layers, and three Transformer encoder layers (see Figure 3a). Specifically, the convolutional layer has a kernel size of 3, a stride of 1, and 64 filters, designed to extract localized audio patterns. Each bidirectional LSTM layer comprises 256 units per direction, capturing sequential dependencies in both forward and backward directions. The Transformer encoder layers each have 8 attention heads and a hidden dimension of 512, enabling long-range contextual understanding through self-attention mechanisms. The convolutional layer captures localized audio patterns, such as changes in pitch or energy. The bidirectional LSTMs model sequential dependencies in both forward and backward directions, enabling the network to understand the temporal flow of audio signals. Finally, the Transformer layers provide long-range contextual understanding through self-attention, allowing the network to focus on key parts of the sequence relevant to emotion recognition. Together, these components transform the low-level LLDs into context-rich audio representations that significantly improve the contribution of the audio modality in multi-modal fusion tasks.

\subsubsection{Textual Features Extraction and Context Network}


\textbf{Textual Features Extraction:} 
Previous studies have established that text is the primary modality in conversational emotion recognition \citep{calvo2010affect, singh2022systematic, cai2023emotion}. However, incorporating a secondary modality alongside text is often challenging due to the inherent differences between modalities (see Figure 1). We aim to evaluate the ability of our proposed HBAF model to bridge and integrate information across different levels of abstraction, achieving our primary goal of exploring interactions between diverse modalities. 

Specifically, the textual representation includes two types of features for each utterance: contextual features and commonsense features. Contextual features, extracted using RoBERTa, capture the semantic relationships and linguistic context within the text, enabling the model to understand nuances based on surrounding words and phrases \citep{liu2019roberta}. On the other hand, commonsense features, derived from COMET, encode background knowledge and inferential reasoning, enriching the text representation with implicit understanding beyond the literal text \citep{sap2019atomic}. Together, these features form a comprehensive high-level text representation. RoBERTa excels at extracting contextual features due to its pretraining on large corpora using masked language modeling, which enables it to model word dependencies and contextual meanings. Similarly, COMET specializes in extracting commonsense features through its training on structured knowledge graphs, allowing it to infer hidden relationships and provide reasoning capabilities.
\begin{figure}[!t]
\centering
\includegraphics[width=1.0
\linewidth]{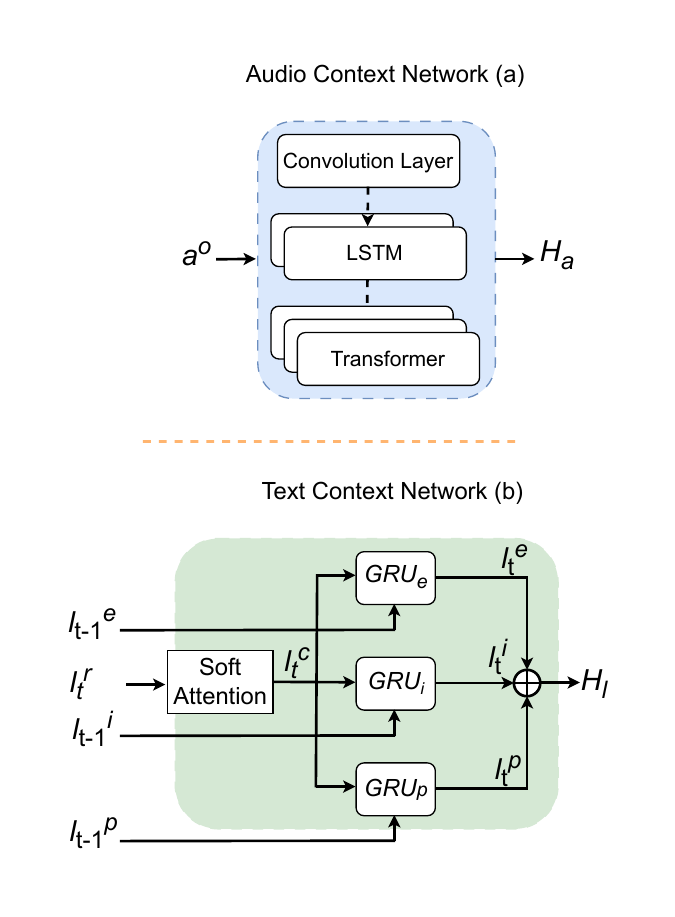}
\caption{Update scheme of the audio context network (a) and text context network (b).}
\label{fig:res2}
\end{figure}

For contextual features, we use the pre-trained RoBERTa Large Model as our language model to extract contextual representations \citep{liu2019roberta}. Each utterance is first passed through the pre-trained RoBERTa model to obtain the outputs from the final four layers, which are then averaged to generate a independent context utterance feature vector ($l^{r}$) with a dimension of 1024. For commonsense features, we use the COMET model to extract commonsense knowledge for emotion prediction \citep{sap2019atomic}. Each utterance is processed by the COMET model to produce three distinct commonsense relations: external state ($l^{e}$), internal state ($l^{i}$), and purpose state ($l^{p}$) \citep{sap2019atomic}. These three relations capture the external, internal, and purposeful states, each with a dimension of 1024, in dynamic conversations. The RoBERTa and COMET features are then passed through a linear layer to generate high-level text features with a dimension of 512.

\textbf{Textual Context Network:} To enhance the contribution of contextual information to the utterance at the time index $t$ , we take into account the independent context utterance feature vectors of the previous utterances using soft attention. The resulting attentive contextual representation $l^{c}_t$ is computed as follows: 

\begin{equation}
    \displaystyle
    u_i \, = \, tanh(W_s l_i^r + b_s), \quad 1 \leq i \leq t - 1
\end{equation}
\begin{equation}
    \displaystyle
    \alpha_i \, = \, \frac{exp(u_i^\mathsf{T})}{\sum_{i=1}^{t-1}exp(u_i^\mathsf{T})}
\end{equation}
\begin{equation}
    \displaystyle
    l_t^c \, = \, \sum_{i=1}^{t - 1} \alpha_i l_i 
\end{equation}

where the parameters $W_{s}$ and $b_s$ are learnable weights and biases.

We combine the attentive contextual representation $l^{c}_t$ with three different commonsense relations using gated recurrent units (\emph{GRUs}) to model the external state ($GRU_{e}$), internal state ($GRU_{i}$), and purpose state ($GRU_{p}$) as follows (see Figure 3b). Each \emph{GRUs} comprises 512 units. \textbf{External state} $GRU_{e}$ aims to capture the participants' external state. At time step $t$, the external state is updated based on the previous external state $l^e_{t-1}$ and the attentive contextual representation $l^c_{t}$. The external state at time step $t$, $l^e_{t}$, can be computed as follows:
\begin{equation}
\displaystyle
    l^e_{t} \, = \, GRU_{e}(l^e_{t-1} \, , \, l^c_{t})
\end{equation}

\textbf{Internal state} $GRU_{i}$ models the internal state of the participants. The previous internal state of the person $l^i_{t-1}$, and the attentive contextual representation $l^c_{t}$ are the inputs to internal state $GRU_{i}$ at timestamp \emph{t}:
\begin{equation}
    \displaystyle
    l^i_{t} \, = \, GRU_{i} (l^i_{t-1} \, , \, l^c_{t})
\end{equation}

\textbf{Purpose state} $GRU_{p}$ carries out the purpose state of the participants. For time step \emph{t}, the purpose state $l^p_{t}$ is updated by taking into account the previous purpose state $l^p_{t-1}$, and the attentive contextual representation $l^c_{t}$. 
\begin{equation}
    \displaystyle
    l^p_{t} \, = \, GRU_{p} (l^p_{t-1} \, , \, l^c_{t})
\end{equation}

Finally, we concatenate the external state $l^e$, internal state $l^i$ and purpose state $l^p$ to form the high-level text representation ($H_l$).

\subsection{Multi-modal Fusion Module}
\label{ssec:subhead}
To effectively learn intra- and inter-modal interactions between low-level audio representation and high-level text representation, we design a multi-modal fusion module which mainly consists of bimodal attention network, dynamic filter gate and residual connection (see Figure 4).  

\subsubsection{Bimodal Attention Network}
The bimodal attention network introduces parallel self-attention and cross-attention mechanisms to learn intra- and inter-modal interactions between low-level audio representation and high-level text representation. Specifically, self-attention operates within the same modality to capture intra-modal information. In this process, the target utterance acts as the query, while the contextual utterances serve as the key, enabling each utterance to determine how much information should be activated from the surrounding contextual utterances.

We estimate the associations within the same modality in a self-attentive manner using scaled dot-attention \citep{vaswani2017attention}. The representations of modality $m$, denoted as $H_m$ where $m$ represents either audio ($a$) or text ($l$), is projected into the query matrix ($Q_m$), key matrix ($K_m$), and value matrix ($V_m$) through linear projections without bias:
\begin{equation} 
    \zeta H_m = \text{softmax}(Q_m K_m^\mathsf{T}/\sqrt{d})V_m
\end{equation} 
where the query ($Q_m$), key ($K_m$), and value ($V_m$) can be interpreted as representations of modality $m$ in different projection spaces. $\zeta H_m$ is the propagated information within modality $m$, $m \in \{a, l\}$. 

\begin{figure}[!t]
\centering
\includegraphics[width=1.0
\linewidth]{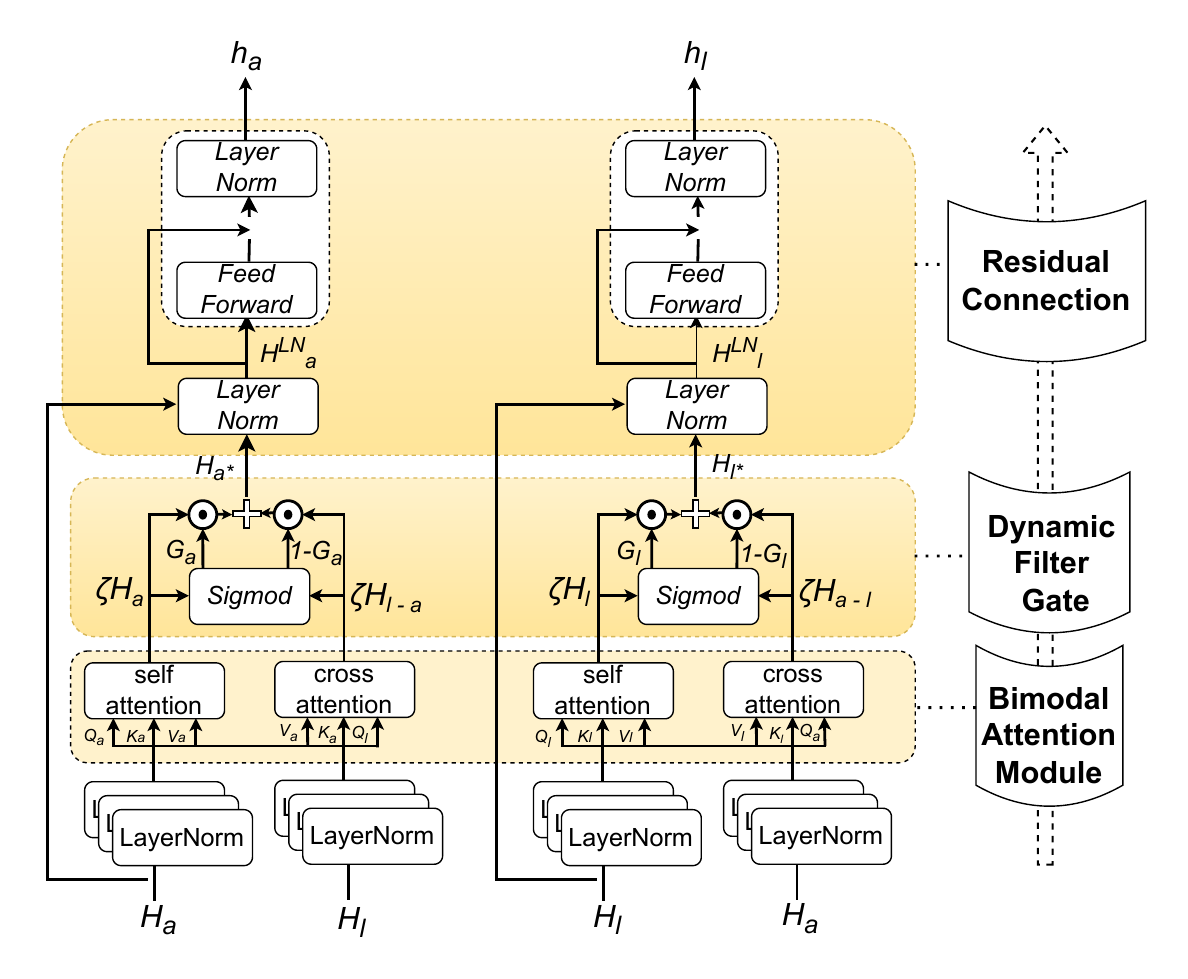}
\caption{Update scheme of the multi-modal fusion module.}
\end{figure}

Parallel to self-attention module, the cross-attention module is used to learn the inter-modal interactions between audio and text modalities. The cross-attention mechanism shares a similar principle with the self-attention mechanism. The only difference is that the query, key and value are from the different modality. The cross-attention module can be described as: 
\begin{equation} 
    \zeta H_{a-l} \, = \, softmax(Q_a K^\mathsf{T}_l/\sqrt{d})V_l
\end{equation} 


\begin{equation} 
    \zeta H_{l-a} = \text{softmax}(Q_l K^\mathsf{T}_a/\sqrt{d})V_a
\end{equation} 

where $\zeta H_{a-l}$ and $\zeta H_{l-a}$ represent the propagated information from audio to text and text to audio, respectively. 

\subsubsection{Dynamic Filter Gate}
Each modality contributes differently to the emotional expression of an utterance, necessitating a mechanism to balance and regulate their interactions during the multi-modal fusion process. To address this, we employ the dynamic filter gate to selectively filter out incorrect or irrelevant cross-modal relations, ensuring that the multi-modal fusion focuses on meaningful and complementary information. The dynamic filter gate operates as follows:

\begin{itemize}
    \item \textbf{Gate Calculation:} For each modality, a gating mechanism is defined to compute the importance of the modality's own information and the cross-modal information from the other modality:
    \begin{equation}
        G_a = \text{sigmoid}\left(\zeta H_{a} \cdot W_{sa} + \zeta H_{l - a} \cdot W_{ca} + b_a\right)
    \end{equation}
   where the parameters $W_{sa}$, $W_{ca}$, and $b_a$ are learnable weights and biases used to model the contribution of self-modal and cross-modal features. The sigmoid function ensures that the gate values are scaled between 0 and 1, controlling the contribution of each component.

    \item \textbf{Modality Refinement:} Once the gate values ($G_a$) are computed, the representation of the modality $H_{a \ast}$ is refined as a weighted combination of its own features and cross-modal features:
    \begin{equation}
        H_{a \ast} = G_a \odot \zeta H_{a} + (1 - G_a) \odot \zeta H_{l - a}
    \end{equation}
    Here, the Hadamard product ($\odot$) is used to apply the gate values to the respective components, ensuring that the fusion process is dynamically weighted based on the gate outputs.

    \item \textbf{Cross-Modal Integration:} The same process is applied to the complementary modality $l$:
    \begin{equation}
        G_l = \text{sigmoid}\left(\zeta H_{l} \cdot W_{sl} + \zeta H_{a - l} \cdot W_{cl} + b_l\right)
    \end{equation}
    \begin{equation}
        H_{l \ast} = G_l \odot \zeta H_{l} + (1 - G_l) \odot \zeta H_{a - l}
    \end{equation}

  where the parameters $W_{sa}$, $W_{ca}$, $W_{sl}$, $W_{cl}$, $b_a$, and $b_l$ are trainable, enabling the model to learn optimal weights for the contributions of self-modal and cross-modal features. The dimensions of $h_{a*}$ and $h_{l*}$ are both 512.
\end{itemize}

Unlike static fusion methods \citep{hazarika2018conversational, wang2022m2r2, shou2024adversarial}, the dynamic filter gate enables the model to adaptively weigh and combine modality-specific and cross-modal features, ensuring robustness against noise and irrelevant information from any modality. The two parallel flows in Fig.~4 represent the processing of two distinct modalities (e.g., audio and text), with each flow preserving its modality-specific features while interacting with the other through cross-modal attention. This design allows the dynamic filter gate to selectively balance the contributions of self-modal features and cross-modal interactions for each modality. 

The dynamic filter gate computes gating values for each modality using a sigmoid function, which evaluates the relative importance of self-modal and cross-modal features. It is described as ``dynamic" because it adjusts these weights for each input, effectively filtering out noisy or irrelevant information and retaining only meaningful and complementary features during the fusion process. By explicitly modeling the dependencies between modalities, this mechanism refines their interactions, resulting in a more accurate, robust, and context-aware multi-modal representation.

\subsubsection{Residual Connection}
To further enhance the representation capacity, we feed the dynamic filter gate features ($H_{a \ast}$ ,  $H_{l \ast}$) into a residual layer to obtain the final interactive information between the audio and text modalities. Specifically, the residual connection consists of a feed forward layer and a normalization layer.

\begin{equation} 
H^{L N}_a \, = \, LayerNorm(H_a  \,\, {\oplus} \, \, H_{a \ast}) 
\end{equation} 

\begin{align} 
h_{a} & \, = \, AddNorm(H^{L N}_a  \,\, {\oplus} \, \,  FeedForward(H^{L N}_a))
\end{align} 

\begin{align} 
H^{L N}_l & \, = \, LayerNorm(H_l \,\, {\oplus} \, \, H_{l \ast})
\end{align}

\begin{align}
h_{l} & \, = \, AddNorm(H^{L N}_l  \,\, {\oplus} \, \,  FeedForward(H^{L N}_l))
\end{align}

We denote the final interactive information between the audio and text modality as $h_a$ and $h_l$, respectively, where $\oplus$ represents the concatenation operation. The dimensions of $h_a$ and $h_l$ are both 512.

\subsection {Inter-modal Contrastive Learning Module}
The traditional approach to contrastive learning relies on self-supervised contrastive learning \citep{gutmann2010noise, radford2021learning}. This method centers on the principle of pulling an anchor and its positive sample closer together in the feature space while pushing the anchor away from negative samples. By doing so, the model learns robust feature representations that effectively capture meaningful patterns, enhancing its ability to distinguish between similar and dissimilar data points.

In this work, we extend the concept of contrastive learning by introducing inter-modality contrastive learning, designed to capture the complex interactions between audio and text modalities (see Figure 5). Unlike traditional approaches that focus on a single modality, our method leverages both the individual characteristics of each modality and their combined representation. To achieve this, we employ two types of contrastive losses that operate on the encoded uni-modal representations ($h_a$ for audio and $h_l$ for text) and the multi-modal fusion representation ($h_m$) during training. These losses ensure that the fused multi-modal representation effectively integrates information while preserving inter-modal relationships.

The first loss, Absolute Inter-Modal Loss, ensures alignment between the uni-modal and multi-modal representations, maintaining consistency across modalities. The second loss, Relative Inter-Modal Loss, captures the subtle dynamics within and between modalities, respecting their inherent differences while reinforcing their complementary nature. Together, these losses enable the model to learn both inter-modal and intra-modal relationships, enhancing its ability to process and interpret multi-modal data. This novel framework represents a significant advancement, particularly in multi-modal emotion detection, by achieving a deeper understanding of the relationships between audio and text modalities. The proposed contrastive learning strategies are as follows:

\begin{figure*}[!t]
\centering
\includegraphics[width=1.0
\linewidth]{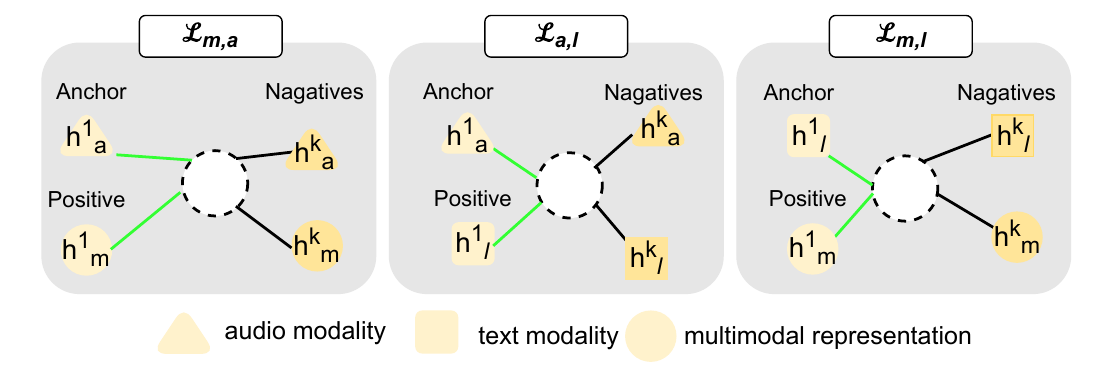}
\caption{Update scheme of the inter-modal contrastive learning module:the absolute inter-modal loss: $\mathcal{L}_{m \, , \ a}$ and $\mathcal{L}_{a \, , \ l}$ and the relative-modal loss: $\mathcal{L}_{m \, , \ a}$.}
\label{fig:res_duplicate}
\end{figure*}

\textbf{The absolute inter-modal loss} focuses on capturing the interaction between the encoded uni-modal representations ($h_a$ and $h_l$) and the encoded multi-modal fusion representation ($h_m$). In this setup, $h_a$ and $h_l$ are concatenated to form the multi-modal representation $h_m$, which serves as the anchor. The audio and text representations from the same sample act as its augmented versions. Each mini-batch is constructed with \emph{K} samples, where each sample consists of an audio representation, a text representation, and a multi-modal fusion representation. For each anchor, a batch of randomly sampled pairs includes two positive pairs and 2\emph{K} negative pairs. Positive pairs are formed by pairing the multi-modal fusion representation with its corresponding audio and text representations from the same sample. Negative pairs are generated by pairing the multi-modal fusion representation with the audio and text representations from different samples. In this case, the absolute inter-modality loss can be formulated as:

\begin{equation}
\mathcal{L}_{a \, , \ m} = - \log \frac{\exp{\left(\left\langle h_a^i, h_m^i \right\rangle  / \tau\right)}} {\exp{\left(\left\langle h_a^i, h_m^i \right\rangle  / \tau\right)} +{\sum_{k=1}^{K} \exp{\left(\left\langle h_a^i, h_m^k \right\rangle  / \tau \right)}}
}
\end{equation}

\begin{equation}
\mathcal{L}_{l \, , \ m} = - \log \frac{\exp{\left(\left\langle h_l^i, h_m^i \right\rangle  / \tau\right)}} {\exp{\left(\left\langle h_l^i, h_m^i \right\rangle  / \tau\right)} +{\sum_{k=1}^{K} \exp{\left(\left\langle h_l^i, h_m^k \right\rangle  / \tau \right)}}
}
\end{equation}
where $\left\langle , \right\rangle$ is cosine similarity. $\mathcal{L}_{a \, , \ m}$, and $\mathcal{L}_{l \, , \ m}$  represent the absolute inter-modal loss of bimodal-audio and bimodal-text, respectively. $\tau$ refers to the temperature hyperparameters for scaling. 

\textbf{The relative inter-modal loss} captures the interactions between the encoded audio ($h_a$) and text ($h_l$)modalities. We consider the audio modality representation as the anchor and the text representation from the same sample as its augmented version. For each anchor, a batch of randomly sampled pairs includes one positive pair and \emph{K} - 1 negative pairs. Here, positive pairs include the audio representation paired with its corresponding text representation from the same sample. Negative pairs are formed by pairing the audio representation with text representation from different samples. In this case, the relative inter-modal loss can be formulated as:

\begin{equation}
\mathcal{L}_{a \, , \ l} = - \log \frac{\exp{\left(\left\langle h_a^i, h_l^i \right\rangle  / \tau\right)}} {\exp{\left(\left\langle h_a^i, h_l^i \right\rangle  / \tau\right)} +{\sum_{k=1}^{K - 1} \exp{\left(\left\langle h_a^i, h_l^k \right\rangle  / \tau \right)}}
}
\end{equation}

Finally, the inter-modal loss function is a weighted sum of absolute and relative inter-modal loss, which can be formulated as: 

\begin{equation}
    \mathcal{L}_{inter} \, = \, \lambda_1 \mathcal{L}_{a \, , \ m} \, + \ \lambda_2 \mathcal{L}_{l \, , \ m} \, + \ \lambda_3 \mathcal{L}_{a \, , \ l}
\end{equation}
where $\lambda_1$, $\lambda_2$, and $\lambda_3$ are hyperparameters to constrain the contributions of the different inter-modal contrastive loss. 

\subsection{Emotion Classification}
The multi-modal representation $h_m$ is passed through a fully connected layer to predict the confidence score $p_c$ for the final emotional state. For the classification task, the standard cross-entropy loss is defined as follows:

\begin{equation}
\mathcal{L}_{ce} = - \sum_{c=1}^{C} y_c \log(p_c)
\end{equation}

where $c$ represents the emotion categories, and $y_c$ indicates the presence based on the ground-truth label.

\subsection{Training}
To train the model, we combine the cross-entropy loss and inter-modal contrastive loss as the overall loss function: 
\begin{equation}
    \mathcal{L} \, = \, \mathcal{L}_{ce} \, + \, \mu \mathcal{L}_{inter} 
\end{equation}
where $\mu$ is hyperparameter to  constrain the contribution of the inter-modal contrastive loss. 

Our experiments indicated that the proposed HBAF model performed best when $\mu$ was set to 0.2. We demonstrate that the paradigm for multi-modal emotion recognition relies on supervised learning, supplemented by self-supervised learning, to effectively and comprehensively achieve emotion understanding. This approach aligns with trends in classification tasks, where supervised learning benefits from labeled data for precise predictions. Self-supervised learning complements this by extracting rich features from unlabeled data, which not only enhances model robustness but also mitigates potential errors introduced by noisy or inconsistent labels in supervised datasets.

\section{Experiments}
\label{sec:intro}
\subsection{Dataset and Metrics}
\label{ssec:subhead}
We evaluated our proposed HBAF method on the two benchmark datasets: MELD \citep{poria2018meld} and IEMOCAP \citep{busso2008iemocap}. Both these two commonly used public datasets are multi-modal, containing audio, text and video modality for every utterance. Table 1 shows the distribution of training and test samples for both two datasets. 

\begin{itemize}
\item MELD is a multi-modal and multi-party dataset for conversational emotion recognition \citep{poria2018meld}. It consists of 13,708 utterances in 1,433 dialogues collecting  from the Friends TV shows. Each utterance is labeled with one in seven emotions: anger, joy, sadness, neutral, disgust, fear and surprise. 

\item IEMOCAP contains videos of dyadic conversations of ten speakers, spanning 12.46 hours \citep{busso2008iemocap}. Each utterance is annotated using the following discrete categories: happy, sad, neutral, angry, excited, and frustrated. 

\end{itemize}

Due to the natural imbalance across various emotions in the dataset, we evaluated the performance of multi-modal emotion recognition tasks using the weighted F1-score, which weights the F1-score for each class based on the proportion of its samples.










\begin{table}[t]
\centering
\footnotesize 
\caption{Statistics of two benchmark datasets: MELD and IEMOCAP}

\begin{tabular}{lcc|cc}
\toprule
\textbf{Dataset} & \multicolumn{2}{c}{\# Conversations} & \multicolumn{2}{c}{\# Utterances} \\
                 & \textbf{train \& val} & \textbf{test} & \textbf{train \& val} & \textbf{test} \\
\midrule
IEMOCAP & 120 & 31 & 5810 & 1623 \\
MELD             & 1153 & 280 & 11098 & 2610 \\
\bottomrule
\end{tabular}
\end{table}

\subsection{Baselines and State-of-the-Art}
\label{ssec:subhead}
In order to comprehensively evaluate the performance of proposed method, we compared the obtained performance with that of the Bidirectional Contextual LSTM (bc-LSTM) baseline \citep{poria2018meld}. This baseline system leveraged an utterance-level LSTM to model context-aware utterance representations from the surrounding utterances. Furthermore, we also compared the proposed framework to various existing state-of-the-art methods. We ran their publicly available models on the two benchmark datasets. The experimental settings, model weights, and datasets were kept consistent across all methods.

\begin{figure*}[htb]
\begin{minipage}[b]{1.0\linewidth}
  \centering
  \centering\includegraphics[width=16cm]{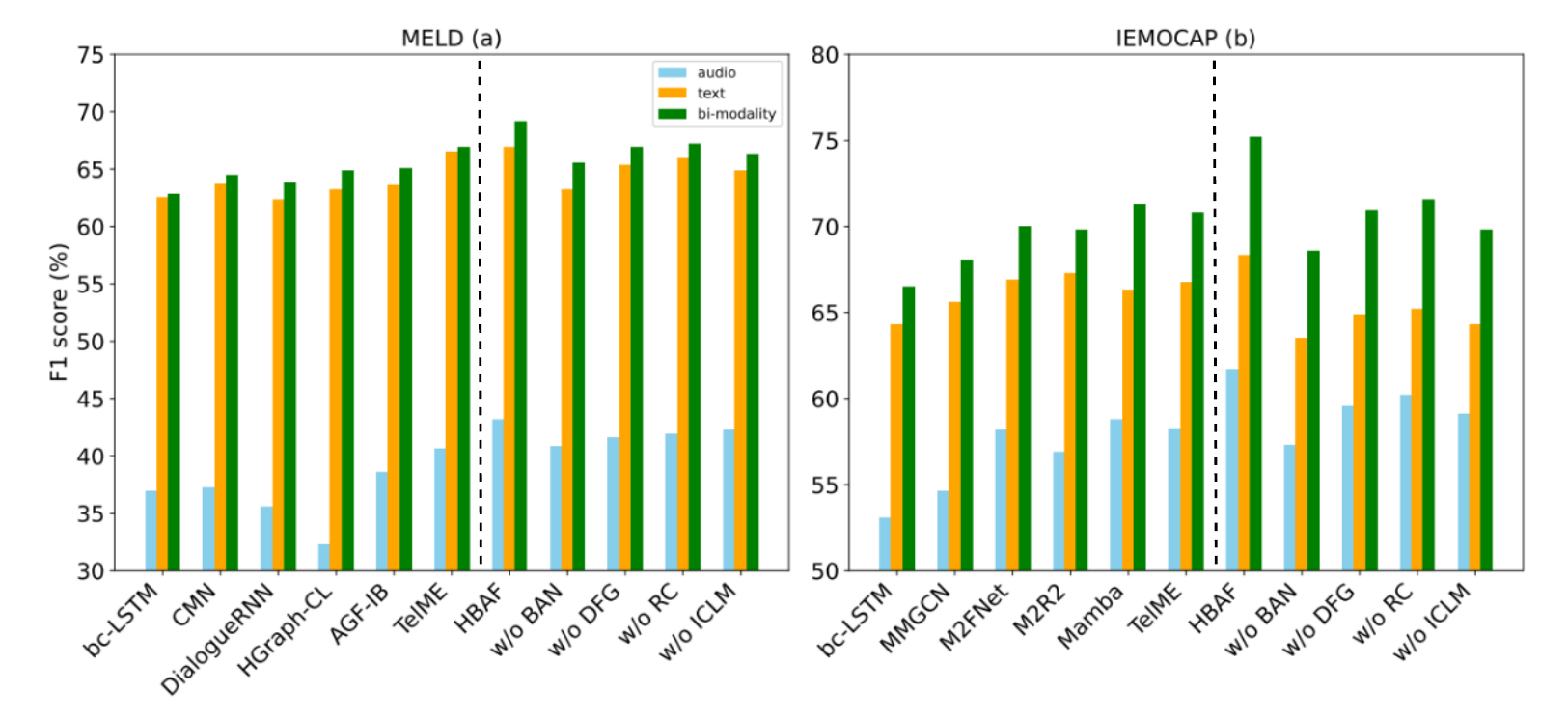}
\end{minipage}
\caption{\small Overall performance comparison among the proposed HBAF method, state-of-the-art methods, and baseline approaches on the MELD (a) and IEMOCAP (b) datasets.}
\label{fig:res}
\end{figure*}

On the MELD dataset, the following systems were used for comparison:
\begin{itemize}
\item CMN \citep{hazarika2018conversational} leveraged two speaker-sensitive \emph{GRUs} to learn contextual utterance combining dialogue history information. These contexts were served as memories to help the prediction of an incoming utterance. 

\item DialogueRNN \citep{majumder2019dialoguernn} deployed three different \emph{GRUs} to track speaker’s states and sequential information in dialogues, and learn inter-modal interactions for conversational emotion recognition.

\item HGraph-CL \citep{majumder2019dialoguernn} used hierarchical graph contrastive learning to model intra- and inter-modal relations for precisely learning sentiment content. 

\item AGF-IB \citep{shou2024adversarial} introduced graph contrastive representation learning to capture intra-modal and inter-modal complementary semantic information and learn intra-class and inter-class boundary information of emotion categories.

\item TelME1 \citep{yun2024telme} incorporated cross-modal knowledge to transfer information from a language model (acting as the teacher) to non-verbal modalities (the students), thereby optimizing the performance of the weaker modalities. 
\end{itemize}

On the IEMOCAP dataset, the following systems were used for comparison:

\begin{itemize}
\item MMGCN \citep{hu-etal-2021-mmgcn} modeled multi-modal dependencies and learned inter-modal interaction for predicting emotions.  

\item M2FNet \citep{Chudasama_2022_CVPR} employed a multi-head attention-based fusion mechanism to combine emotion-rich latent representations of emotion-relevant features from visual, audio, and text modality and learn intra- and inter-modal relationships. 

\item M2R2 \citep{wang2022m2r2} utilized \emph{RNN} to handle incomplete utterances by spreading dependency on all conversation parties. 

\item Mamba \citep{shou2024revisiting} designed a multi-modal fusion strategy based on probability guidance to maximize the consistency of information between modalities and learned intra- and inter-modal interactions for conversational emotion recognition. 

\item TelME1 \citep{yun2024telme} incorporated cross-modal knowledge transfer, where a language model (acting as the teacher) transfers information to non-verbal modalities (the students), thereby optimizing the performance of the weaker modalities. 

\end{itemize}

\subsection{Model Configuration}
Following \citep{luo2023crossmodal}, we implemented our proposed HBAF method using the Pytorch 1.11.0 framework. The model was trained with Adam optimizer with an initial learning rate of 1e-4 and a batch size of 32. To mitigate overfitting, the network was regularized by L2 norm of the model’s parameters with a weight of 3e-4. The model training was stopped if the validation loss did not decrease for 15 consecutive epochs. 

\section{Results and Discussion}
To verify the effectiveness of our proposed HBAF method, we conduct comparative experiments against previous state-of-the-art baselines. Next, we present uni-modal and bimodal results, analyzing the contribution of each modality, particularly different level audio representation. Following that, we emphasize the importance of uni-modal representation module, multi-modal fusion module and inter-modal contrastive learning module. Finally, we conduct case studies and error analysis.

\subsection{Comparison with State-of-the-art Baselines}
To verify the effectiveness of our proposed HBAF method,  we present our comparative studies against state-of-the-art baseline systems (see Figure 6). HBAF demonstrates superior performance over state-of-the-art baseline systems in terms of weighted F1-score, achieving a 2.2\% improvement over TelME1 on the MELD dataset and a 6.9\% improvement over Mamba on the IEMOCAP dataset. These results demonstrate the superior expressive power and efficacy of incorporating low-level audio representation into the high-level text representation for bimodal speech emotion recognition. Notably, bimodal speech emotion recognition models consistently outperform most single-modality models. This improvement can be attributed to three factors: the consolidation of contextual information, the interaction within multi-modal fusion, and the quality of inter-modal contrastive learning. HBAF achieves an F1-score improvement of 2.3\% and 5.1\% over text-based uni-modal models on the MELD and IEMOCAP datasets, respectively, when compared to our proposed HBAF model relying solely on textual input. This highlights the significant advantage of integrating multi-modal features for enhanced performance.

It is worth noting that the performance gains on the MELD dataset are more subtle compared to the IEMOCAP dataset. HBAF shows the strongest ability to infer emotions like joy, anger, and sadness in the MELD dataset, likely because these emotions are explicitly expressed through both audio and text. However, performance on fear is lower across most models, possibly due to its implicit expression and the limited number of samples.

Upon further analysis, we observed that the dialogues in the MELD dataset are relatively shorter (mostly 5 to 9 utterances) compared to those in the IEMOCAP dataset, which average around 70 utterances per dialogue. Additionally, the MELD dataset is based on real-world scenarios, which contain substantial background noise (e.g., honking, barking). This noise may be introduced into the model, particularly for emotions like fear and frustration, which can be easily masked by noise. Consequently, Our proposed HBAF method achieves better performance on the IEMOCAP dataset. 

\subsection{Importance of Modalities}
\label{ssec:subhead}
We examined the importance of each modality in this section. We chose audio, text, or bi-modality as the input for conversational emotion recognition. Specifically, for the audio modality, we fed different low-level and high-level audio representation into HBAF framework for emotion prediction. We show all uni-modal and bimodal results in Figure 6 and Table 2.

\begin{table}[t]
\centering
\footnotesize 
\caption{Ablation Study of the Audio Context Network on Different Levels of Audio Representation (dark gray color indicates the low-level audio representation). The ACN denotes audio context network. The green arrow indicates an improvement in performance when using ACN compared to without it. The audio input for our proposed HBAF model is  LLDs representation.)}
\begin{tabular}{c c c c c c}
\toprule
\multirow{2}{*}{Representation} & \multicolumn{2}{c}{MELD} & \multicolumn{2}{c}{IEMOCAP} \\ \cmidrule(lr){2-3} \cmidrule(lr){4-5}
                          & w/ ACN & w/o ACN & w/ ACN & w/o ACN \\ 
\midrule
\rowcolor{lightgray} \emph{LLDs} & 0.412 \textcolor[rgb]{0,1,0}{$\uparrow$} & 0.398   & 0.583  \textcolor[rgb]{0,1,0}{$\uparrow$} & 0.567   \\ 
\rowcolor{lightgray} \emph{MFCCs} & 0.346 \textcolor[rgb]{0,1,0}{$\uparrow$} & 0.335   & 0.528  \textcolor[rgb]{0,1,0}{$\uparrow$} & 0.515   \\ 
\emph{VGGish}                 & 0.422  & 0.423   & 0.585  & 0.589   \\ 
\emph{openL3}                   & 0.435  & 0.437   & 0.591  & 0.593   \\ 
\emph{wav2vec}                   & 0.473  & 0.476   & 0.619  & 0.621   \\ 
\emph{data2vec}                  & 0.379  & 0.378   & 0.530  & 0.527   \\ 
\bottomrule
\end{tabular}
\end{table}

For uni-modal average weighted F1-score results (see Figure 6), it should be highlighted that text modality significantly outperforms audio modality on MELD by 23.6\% (HBAF) to 30.9\% (HGraph-CL) and IEMOCAP by 6.2\% (HBAF) to 11.2\% (bc-LSTM). This is plausible for at least two reasons: 1) acoustic information may sometimes confuse emotion recognition task. For instance: “angry” and “joy” may have similar acoustic performances (high volume and pitch) even though they belong to opposite sentiments; and 2) audio signal has a lot of background noise. Thus, the essential problem is how to find the best audio representation due to the complex and variation of audio features. What's more, the word-level lexical features contains a wealth of information, such as topic, intent, instance, and so on. The content is important to release the nature and flow of the emotional dynamics of participants in conversations. 

We hypothesize that our proposed HBAF method effectively bridges the heterogeneous modality gap between different levels of modality representation by leveraging its key components, particularly the Audio Context Network (ACN). The ACN plays a vital role in aligning audio features with other modalities, ensuring smoother integration and improving the overall fusion process. To verify the effectiveness of HBAF, we maintained the same high-level text representation and introduced different levels of audio representation (see Table 2). Specifically, we explored various audio representations, including handcrafted feature sets (e.g., \emph{LLDs} \citep{eyben2010opensmile}), spectral features (e.g., \emph{MFCCs} \citep{davis1980comparison}), and pre-trained audio representations (e.g., \emph{VGGish}, \emph{openL3}, \emph{wav2vec}, \emph{data2vec}) \citep{liu2022audio}. We find that the audio context module is effective for low-level audio representation such as LLDs and MFCC, but not for high-level representation like VGGish, openL3, wav2vec,  data2vec (see Table2). We suspect this is because the low-level audio representation lack contextual content, whereas high-level audio representation already contain contextual content obtained through deep pre-trained models. 

However, emotion recognition in conversations sometimes is contextual dependent to its audio, rather than text, and relies on features like voice tone, pitch, speed, and other audio information to infer what the emotion of a participant in conversations. For example, a single word ``okay'' is uncertainty and ambiguous, and it can express a variety of emotions in different situations. After fusing corresponding audio features, it is not difficult to infer the sentiment of this short utterance is negative. In general, integrating information from multiple modalities tends to achieve better performance than those under single-modal collection. 

\subsection{Ablation Study}
To validate the effectiveness and reasonableness of each component of our proposed HBAF method, we carefully ablated the model's audio context network, multi-modal fusion module, and inter-modal contrastive learning module (see Figure 6). We observe the full version of HBAF achieves the best performance on the MELD and IEMOCAP datasets. We place particular emphasis on the removal of either the audio contextual network, multi-modal fusion module, or inter-modal contrastive learning module, each of which adversely affected the model's results. This indicates that the multi-modal fusion module comprising of bimodal attention network, dynamic filter gate and residual connection is crucial for fully learning intra- and inter-modal interaction for bimodal speech emotion recognition. 


\subsubsection{Effect of the Audio Context Network} 
We observe that the absence of the audio contextual module (ACN) for low-level audio representations causes a noticeable performance decrease (see Table 2). Specifically, on the MELD dataset, the performance drops by 1.4\% for \emph{LLDs} and 1.1\% for \emph{MFCCs}, while on the IEMOCAP dataset, the performance decreases by 1.6\% for \emph{LLDs} and 1.3\% for \emph{MFCCs}. This highlights the critical role of the ACN in enriching low-level audio features by incorporating contextual information, which substantially enhances the model's performance across both datasets. However, the effect of the audio context network on high-level audio representation is very weak and may even have negative side effects. This result can be explained by the fact that the audio contextual network introduces contextual information to low-level audio representation learning, promoting semantic alignment for bridging heterogeneous modality gap. 
 
By leveraging different level audio and textual representations, recent fusion models are geared toward directly modeling intra- and inter-modal interactions \citep{hazarika2018conversational, majumder2019dialoguernn, shou2024revisiting, yun2024telme}. However, such approaches suffer from a heterogeneous modality gap in multi-modal fusion interactions, particularly concerning the information discrepancy between low-level and high-level uni-modal representations. This is because audio and text modalities may inevitably map into different semantic spaces when they are input into different pre-trained models to generate uni-modal representations. Specifically, a sequence of words is processed by a transformer-based model to produce high-level text representation. In contrast, raw audio data is conducted simple statistical analysis to generate low-level audio representation.

\begin{figure*}[!t]
\centering
\includegraphics[width=0.8
\linewidth]{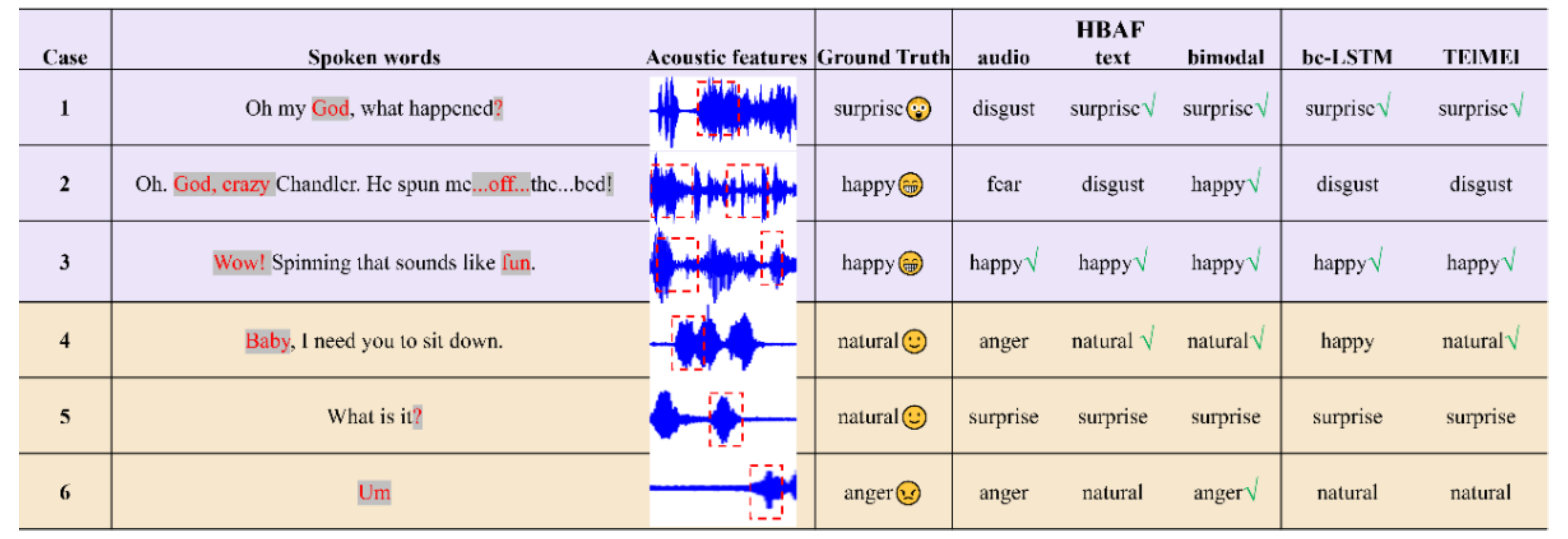}
\caption{Input of audio, text and bimodality and prediction with HBAF method, bc-LSTM and TEIMEI on MELD and IEMOCAP datasets in our case study.}
\label{fig:res}
\end{figure*}

We contend that directly fusing audio and text representations at different levels into a common multi-modal embedding space is suboptimal, as it may lead to heterogeneous conflicts during the fusion stage. In other words, the transformer-based model imparts contextual information to the text modality using both linear and nonlinear structures, but the low-level audio representation do not undergo these operations. However, contextual information is a crucial factor influencing emotion, and when context is incorporated into uni-modal representations, we can infer more emotional states. To bridge this heterogeneous gap, a audio context module comprising of a one-dimensional convolutional layer, two bidirectional long short-term memory layers, and three transformer encoder layers unleashes the full contextual power of complex feature inter-correlations by aligning multi-modal interactions. These observations underscore the necessity and effectiveness of exploring context semantic alignment before fusing multi-modal representations.

\subsubsection{Effect of the Multi-modal Fusion Module}
The multi-modal fusion  module is designed to leverage the dynamic attention weight to learn intra- and inter-modal interactions between the low-level audio representation and the high-level text representation. As shown in Figure 5, the multi-modal fusion module contributes to a performance improvement of 3.6\% on the MELD dataset and 6.6\% on the IEMOCAP dataset, compared to our proposed HBAF model without the multi-modal fusion module. Specifically, the performance decline follows a consistent trend when different components are removed from the model: bimodal attention network $>$ inter-modal contrastive learning module $>$ dynamic filter network $>$ residual network (see Figure 5). Combining information from self- and cross-attention is intuitively beneficial, as they complement and enhance each other, thus providing richer emotion-relevant information for bimodal speech emotion recognition. 

\begin{figure*}
\centering
\includegraphics[width=0.8
\linewidth]{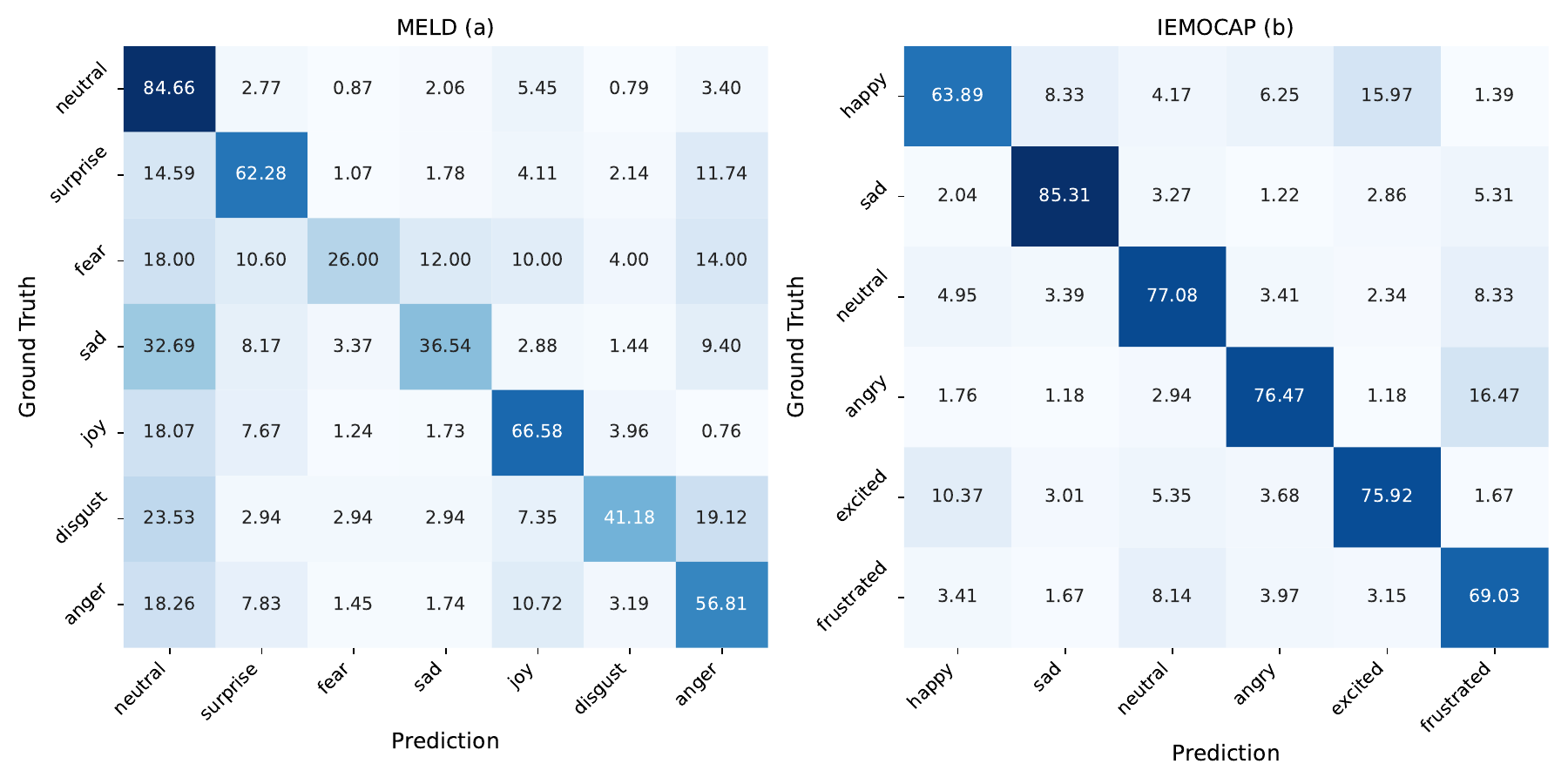}
\caption{ Visualization confusion matrices on the test set of MELD (a) and IEMOCAP (b) }
\label{fig:res}
\end{figure*}

In the ablation study of the multi-modal fusion module, we attribute the performance improvement to the multi-modal fusion module's explicit modeling of intra- and inter-modal interactions within and between audio and text, as well as its ability to capture long-term contextual information. By introducing dynamic self- and cross-modal attentive weights to uni-modal features, we enhance the uni-modal representations to build robust and effective multi-modal representation. We speculate that the multi-modal fusion module can leverage heterogeneous knowledge in a new high-dimensional space to capture more detailed information embedded in each modality, while adaptively fusing implicit complementary content to enhance interactions and correlations. This multi-modal fusion module improves performance by encouraging the proposed model to explore complementary information and interactions between the audio and text modalities during the dynamic interaction process.

We also observe improved performance from the proposed model when utilizing the dynamic filter gate. The dynamic filter gate filters information generated by incorrect cross-modal interactions, selectively removing or adding data to refine the complete emotion-relevant representation of multi-modal inter-correlations. Additionally, the residual connection offers an efficient method to explore heterogeneous spaces, learn modality-specific information, and facilitate synchronization.

\subsubsection{Effect of the Inter-modal Contrastive Learning Module}
Inter-modal contrastive learning aims to fully explore inter-modal interactions.When the inter-modal contrastive learning module is removed, the performance of HBAF drops by 2.9\% on the MELD dataset and 5.4\% on the IEMOCAP dataset, compared to the full model with the module included. This suggests that the inter-modal contrastive learning module facilitates emotional interactions between modalities, leading to improved emotion recognition. Multi-modal interaction based on contrastive learning enables uni-modal interactions to sufficiently capture multi-modal complementary. This is because supervised models may inevitably lose some distinct characteristics of each modality during multi-modal fusion. Under the guidance of self-supervised signals, inter-modal contrastive learning module leverages both absolute and relative inter-modal interactions to fully exploit inter-modal characteristics and enhance different level modality interaction. 

\subsection{Case Studies}
To qualitatively validate the effectiveness of our proposed HBAF, we visualize several typical examples of MELD and IEMOCAP in Figure 7, respectively. Specially, we compare the difference between audio, text and bimodal for conversational emotion recognition. HBAF provides closer scores to the ground truths than uni-modal audio and text, owing to fully capture intra- and inter-modal interactions between audio and text modality. The cases indicate that HBAF can effectively integrate non-verbal modalities with verbal modality. Combining information from audio and text modalities provides richer, emotion-relevant insights by allowing modalities to complement or augment one another. In addition, we also analyze the wrong examples to indicate the shortcomings of single modal for emotion prediction. As shown in Figure 6, HBAF achieves incorrect classification results on cases 5. We analyze the content of these cases. For example, the single word `Um' is ambiguous and can convey positive, neutral, or negative emotions depending on the context. It is challenging to determine the associated emotion based solely textual utterance. The text modality of the case 6 misclassify anger as neutral. Therefore, the bias of the text modality leads to wrong classification results. However, audio modality contains high pitch and tense, shaky voice, the sentiment behind the single word becomes easier to interpret as negative. 

\subsection{Error Analysis}
\label{ssec:subhead}
We visualize the confusion matrices of the test set in Figure 8. For the MELD dataset, our proposed HBAF method performs well across most emotion categories, including neutral, joy, sadness, and surprise. However, emotions like disgust and fear are under-represented in MELD. For the IEMOCAP dataset, HBAF excels in recognizing sadness, excitement, and frustration. We also note that the neutral emotion can be confused with excitement. 

By inferring pre-defined emotions in conversations, we demonstrates that the errors made by the HBAF are mainly caused by the following aspects. First, emotion is a subjective concept with involved uncertainty, resulting in human bias introduced during the labeling of the utterances. The emotion felt by the speaker and perceived by human annotators may have different stances. HBAF sometimes got confused and miss-classified similar or close emotions such as disgust and sad. Secondly, the MELD dataset is highly imbalanced. In particular, the system could not deal with minority classes very well, such as fear and disgust). Many emotions are overwhelmingly predicted as neutral for MELD dataset. These phenomena also appear in previous works \citep{cai2023emotion, singh2022systematic, ezzameli2023emotion}. What's more, human express emotion by various modalities such as facial expressions, voice characteristics, linguistic content of verbal communications, and body postures, while HBAF method has only focused on audio and text modality for predicting emotions in conversations. 

\section{Conclusion}
\label{sec:majhead}
In this paper, we proposed a bimodal framework for bimodal speech emotion recognition, ``Heterogeneous Bimodal Attention Fusion (HBAF)’’, as an early attempt to explore the interactions between audio and text modalities at the different representation level. HBAF was built on three key modules: uni-modal representation module, multi-modal fusion module and inter-modal contrastive learning module. Specifically, we incorporated contextual content into low-level audio representations to bridge the heterogeneous multi-modal gap and improve multi-modal fusion. At the heart of HBAF is the introduction of multi-modal fusion module comprising of bimodal attention network, dynamic filter gate and residual connection, which enables effective and efficient exploration of filtering out incorrect cross-modal relationships and learning both intra- and inter-modal interactions between audio and text modality. Furthermore, the inter-modal contrastive learning module enables fully capturing absolute and relative inter-modal interactions towards the understanding of emotions. Experiments demonstrated that our HBAF outperforms state-of-the-art methods. 

\section{Acknowledgment}
We would like to thank the anonymous reviewers for their valuable comments and feedback. Special acknowledgment is given to the Centre for Digital Music at Queen Mary University of London for its support. This research was funded by the China Scholarship Council and Queen Mary University of London.

\bibliographystyle{elsarticle-harv} 
\bibliography{example}






\end{document}